\documentclass[12pt,runningheads]{llncs}
\usepackage{amsmath, amssymb, mathtools, xcolor}
\usepackage[hidelinks]{hyperref}
\usepackage{relsize, abraces, setspace, enumitem}
\usepackage{graphicx, subfigure, caption}
\usepackage{placeins}
\usepackage{orcidlink}
\usepackage[paper=a4paper,left=3cm,right=3cm,top=3cm,bottom=3cm]{geometry}

\newcommand{ \C }{ \mathcal C }
\newcommand{ \D }{ \mathcal D }
\newcommand{ \Z }{ \mathbb{Z} }
\newcommand{ \Beq }{ B^{\textnormal{e}} }
\newcommand{ \Teq }{ T^{\textnormal{e}} }
\newcommand{ \myx }{ \boldsymbol{x} }
\newcommand{ \myy }{ \boldsymbol{y} }
\newcommand{ \myz }{ \boldsymbol{z} }
\newcommand{ \myw }{ \boldsymbol{w} }
\newcommand{ \myu }{ \boldsymbol{u} }
\newcommand{ \myv }{ \boldsymbol{v} }
\newcommand{ \dist }{ d }
\newcommand{ \run }{ \operatorname{run} }

\newcommand{ \bs }[1]{ \boldsymbol{#1} }

\newcommand{\squeezeddots}{\mathinner{\!\ldotp\! \ldotp\! \ldotp}}
\newcommand{ \defeq }{ \coloneqq }
\newcommand{ \myqed }{ \hfill $\blacklozenge$ }
\renewcommand{\qed}{\hfill $ \blacksquare $}

\DeclareGraphicsExtensions{.pdf}

\begin{document}

\title{Bounds on Codes Correcting Adjacent Transpositions}

\author{Mladen Kova\v{c}evi\'{c}\inst{1}\thanks{Corresponding author: \email{kmladen@uns.ac.rs}}\and
Han Mao Kiah\inst{2}  %HMKiah@ntu.edu.sg
\and
Keshav Goyal\inst{3}} %keshav.goyal4@gmail.com

\authorrunning{M. Kova\v{c}evi\'{c}, H. M. Kiah, K. Goyal}

\institute{University of Novi Sad, Serbia
\and
Nanyang Technological University, Singapore
\and
Universit\'{e} C\^{o}te d'Azur, CNRS, I3S lab, France}

\maketitle

\begin{abstract}
We study the problem of correcting pairwise disjoint adjacent transpositions (or swaps) in $q$-ary strings.
Equivalently, the model we assume is the radius-one instance of the so-called $\ell_\infty$-limited permutation channel.
We first study the relevant combinatorial properties of the appropriately defined transposition distance, including center-specific and average ball sizes.
We then derive two lower bounds and one upper bound on the asymptotic rates of optimal codes correcting $t=\tau n$ transpositions.
The first achievability result is a generalized Gilbert--Varshamov bound, while the second follows from a construction of codes correcting all possible patterns of
adjacent transpositions and therefore represents a lower bound on the zero-error capacity of this model.
This construction improves the classical general-alphabet construction for
$3\leqslant q\leqslant 8$ as well as the recent bounds for $q=4,5$.
The upper bound is obtained by a packing argument adjusted to the run-structure of a given code.
To the best of our knowledge, these are the first nonconstant,
$\tau$-dependent lower and upper bounds developed for the pairwise disjoint
$q$-ary model throughout the linear regime.
We also derive asymptotic bounds on the cardinality of optimal codes correcting
$t=\textrm{const}$ pairwise disjoint adjacent transpositions.

\keywords{error correction \and zero-error capacity \and adjacent transposition
\and timing error \and synchronization error \and reordering error \and permutation channel %\and bit shift
\and Gilbert--Varshamov bound\\\\
{\bf MSC 2020:} 94B25, 94B50, 94B65, 68P30, 68R05}
\end{abstract}

\section{Introduction and preliminaries}
\label{sec:intro}

%Local reordering errors represent a natural intermediate model between memoryless symbol errors and unrestricted permutations.
Local reordering errors occur in various communication settings, e.g., as
spelling errors \cite{damerau}, timing and synchronization errors such as
packet reordering in communication networks \cite{bennett} or transpositions
in molecular communications \cite{haselmayr}, intersymbol interference errors
such as peak-shifts in magnetic recording channels \cite{shamai+zehavi} or
charge drift in flash memories \cite{barg+mazumdar}, genome rearrangement
errors \cite{christie} which are of relevance for DNA-based data storage
systems \cite{gabrys}, etc.
In the present paper we focus on one such model where the transmitted strings are potentially
altered by transpositions (or swaps) of adjacent symbols.
More specifically, we assume that all pairs of symbols that are being swapped in the channel are pairwise disjoint and thus that every input symbol may be displaced by at most one position.
This model is also known as the radius-one
$\ell_\infty$-limited permutation channel \cite{langberg}, denoted
$\operatorname{LPC}_\infty(1)$.

In the remainder of this section we precisely describe and discuss the model we are
interested in, as well as another closely related model.
In particular, we introduce a distance function that is appropriate for our
setting in the sense that it gives a characterization of the corresponding
error correction problem (Section~\ref{sec:model}).
We then analyze the effects of adjacent transpositions on $ q $-ary strings and
the cardinalities of appropriately defined balls,
including an exact enumeration of their average sizes (Section~\ref{sec:balls}).
Our main results, presented in Section~\ref{sec:bounds}, are lower and upper
bounds on the largest achievable asymptotic rate of codes correcting
$ t = \tau n $ adjacent transpositions.
The first is a generalized Gilbert--Varshamov bound (Section~\ref{sec:GV}).
The second is a lower bound on the zero-error capacity of our model and follows from a construction of codes correcting all possible patterns of
adjacent transpositions (Section~\ref{sec:ze}).
The upper bound is obtained by taking into account the run-structure of a code and applying an appropriate packing argument (Section~\ref{sec:run-upper}).
Finally, we derive asymptotic bounds on the cardinality of optimal codes
correcting $ t = \textrm{const} $ adjacent transpositions (Section~\ref{sec:codet}).
Section~\ref{sec:conclusion} concludes the paper.

\subsection{Model description and basic definitions}
\label{sec:model}

Throughout the paper, $ q $ is assumed to be an integer greater than or equal to
$ 2 $, and the channel alphabet is denoted by $ \Z_q = \{\mathtt{0, 1, \ldots, q-1}\} $.
The possible inputs and outputs from the channel are the strings from $ \Z_q^n $,
where $ n \geqslant 1 $ is referred to as the block-length.

For a transmitted string $ \myx = x_1 \ldots x_n $, we say that an adjacent
transposition%
\footnote{In the rest of the paper, we shall usually omit the word `adjacent' and
refer to these types of impairments simply as transpositions. No confusion should
arise as this is the only kind of transpositions discussed in the paper.}
at location $ k \in \{1, \ldots, n-1\} $ has occurred in the channel if the string
obtained at the channel output is $ \myy = y_1 \ldots y_n $, where $ y_k = x_{k+1} $,
$ y_{k+1} = x_k $, and $ y_i = x_i $ for all $ i \notin \{k, k+1\} $, i.e., if the
symbols $ x_k $ and $ x_{k+1} $ have been swapped.
If $ \myx $ is altered by $ t \geqslant 1 $ transpositions, it is assumed that these
transpositions affect disjoint pairs of symbols (one could imagine them being applied
simultaneously to $ \myx $).
More precisely, if the transpositions occur at locations $ k_1, \ldots, k_t $, it
is understood that $ k_{j+1} \geqslant k_j + 2 $ for $ j = 1, \ldots, t-1 $.
Consequently, every symbol of $ \myx $ can be displaced by at most one position in
the channel.
Here is an example of a transmitted string $ \myx \in \Z_4^{10} $ and the received
string $ \myy $ obtained after $ t = 3 $ transpositions (the pairs of symbols that
are being swapped are underlined):
\begin{equation}
\label{eq:xy1}
\begin{aligned}
%\label{eq:transp11}
  \myx   \ = \ \  \aunderbrace[l1r]{\mathtt{0 \ 1}} \ &\mathtt{1} \, \aunderbrace[l1r]{\mathtt{3 \ 0}} \, \mathtt{0 \ 2 \ 2} \, \aunderbrace[l1r]{\mathtt{2 \ 1}}  \\
%\label{eq:transp12}
          \hookrightarrow   \ \mathtt{1 \ 0} \ &\mathtt{1 \ 0 \ 3 \ 0 \ 2 \ 2 \ 1 \ 2}  \ = \  \myy .
\end{aligned}
\end{equation}

We denote by $ B(\myx; t) $ the set of all strings that can be obtained by applying
at most $ t $ transpositions, of the kind just described, to the string $ \myx $.
It will be helpful to also introduce special notation $ \Beq(\myx; t) $ for the set
of all strings that can be obtained by applying \emph{exactly} $ t $ transpositions
to $ \myx $ but cannot be obtained by applying fewer than $ t $ transpositions to
$ \myx $.
The set $ B(\myx; t) $ is therefore a disjoint union of the sets $ \Beq(\myx; t') $
over all $ t' \in \{0, 1, \ldots, t\} $,
\begin{align}
\label{eq:BBequnion}
  B(\myx; t) = \bigsqcup_{t'=0}^{t} \Beq(\myx; t') .
\end{align}
Note that $ \Beq(\myx; t) = \varnothing $ for $ n < 2t $ because the largest possible
number of disjoint transpositions in a string of length $ n $ is $ \lfloor n/2 \rfloor $.
Thus, $ B(\myx; t) = B(\myx; \min\{t, \lfloor n/2 \rfloor\}) $.

A nonempty subset of $ \Z_q^n $ is called a code of length $ n $; its elements
are called codewords.
A code $ \C $ is said to be able to correct $ t $ transpositions if the following
condition holds:
\begin{align}
\label{eq:disjoint}
  \forall \myx, \myx' \in \C  \qquad  \myx \neq \myx'  \;\Rightarrow\;  B(\myx; t)  \cap  B(\myx'; t)  =  \varnothing .
\end{align}
In words, any codeword $ \myx \in \C $, after being altered by up to $ t $
transpositions in the channel, can be uniquely recovered from the received string
(which belongs to $ B(\myx; t) $) because no other codeword could have produced
that string.
Let $ M^\star_q(n; t) $ denote the maximum cardinality of a code in $ \Z_q^n $
correcting $ t $ transpositions,
\begin{align}
\label{eq:T}
   M^\star_q(n; t) \defeq \max \big\{ |\C| \colon \C \subseteq \Z_q^n, \, \C\, \textnormal{corrects $ t $ transpositions} \big\} .
\end{align}
We say that $ \C $ is able to correct an arbitrary number of transpositions (or that
it is a zero-error code) if the relation \eqref{eq:disjoint} holds for every
$ t \geqslant 0 $.
Since the largest possible number of disjoint transpositions in a string of length
$ n $ is $ \lfloor n/2 \rfloor $, this is equivalent to writing
\begin{align}
\label{eq:zedisjoint}
  \forall \myx, \myx' \in \C  \qquad  \myx \neq \myx'  \;\Rightarrow\;
	B(\myx; \lfloor n/2 \rfloor)  \cap  B(\myx'; \lfloor n/2 \rfloor)  =  \varnothing .
\end{align}
In other words, the statements that a code corrects $ \lfloor n/2 \rfloor $
transpositions and that it corrects an arbitrary number of transpositions are equivalent.

We define the transposition distance between $ \myx, \myy \in \Z_q^n $ by
\begin{align}
\label{eq:distance}
  \dist(\myx, \myy)  \defeq  \min \big\{ r + s \colon r,s \geqslant 0, \,  B(\myx; r) \cap B(\myy; s) \neq \varnothing \big\} ,
\end{align}
with the convention $ \min \varnothing = \infty $.
In words, we look at all the ways a common string $ \myz $ can be produced by applying
transpositions to both $ \myx $ and $ \myy $, and among them we choose one requiring
the smallest possible cumulative number of transpositions.
That number is denoted by $ \dist(\myx, \myy) $.
If no common string can be produced by applying transpositions to $ \myx $ and $ \myy $,
i.e., if $ B(\myx; r) \cap B(\myy; s) = \varnothing $ for all $ r, s $, we set
$ \dist(\myx, \myy) = \infty $.
Again, since the largest possible number of disjoint transpositions in a string of length
$ n $ is $ \lfloor n/2 \rfloor $, we have
\begin{align}
\label{eq:infdist}
  \dist(\myx, \myy) = \infty  \quad\Leftrightarrow\quad
	B(\myx; \lfloor n/2 \rfloor) \cap B(\myy; \lfloor n/2 \rfloor) = \varnothing .
\end{align}

The ball of radius $ t $ around $ \myx $ with respect to the distance $ \dist(\cdot,\cdot) $
is denoted by
\begin{align}
\label{eq:ball}
  \bar{B}(\myx; t) \defeq \{ \myy \colon \dist(\myx,\myy) \leqslant t \} .
\end{align}
One can think of $ \bar{B}(\myx; t) $ as the set of all strings that can be obtained
through a two-step process: first apply $ r $ transpositions to $ \myx $, and then
apply $ s $ transpositions to the resulting string, for any $ r, s $ satisfying
$ r + s \leqslant t $.
We note that $ B(\myx; t) \subseteq \bar{B}(\myx; t) $, and that this containment
relation is in most cases strict.

The function $ \dist(\myx,\myy) $ is symmetric, non-negative, and equals zero if
and only if $ \myx = \myy $.
However, it does not satisfy the triangle inequality and is therefore not a metric.
E.g., for $ \myx = \mathtt{1 0 0 0} $, $ \myy = \mathtt{0 0 0 1} $, $ \myz = \mathtt{0 0 1 0} $, we have
$ \dist(\myx, \myy) = \infty $, while $ \dist(\myx, \myz) + \dist(\myz, \myy) = 2 + 1 = 3 $.
Nonetheless, it is a convenient way of measuring distance between strings in the
context of the problem currently being studied;
in particular, it gives a sufficient condition
for the error-correction capability of codes.

Define the minimum distance of a code $ \C \subseteq \Z_q^n $ in the usual way:
\begin{align}
\label{eq:mindist}
  \dist_{\min}(\C)  \defeq  \min \big\{ \dist(\myx, \myx') \colon \myx,\myx' \in \C, \, \myx \neq \myx' \big\} .
\end{align}

\begin{proposition}
\label{thm:eccmetric}
Let $ \varnothing \neq \C \subseteq \Z_q^n $.
(a) If $ \dist_{\min}(\C) > 2t $, then $ \C $ corrects $ t $ transpositions;
(b) $ \C $ corrects an arbitrary number of transpositions if and only if
$ \dist_{\min}(\C) = \infty $.
\end{proposition}
\begin{proof}
We first show the contrapositive of (a).
If the code $ \C $ is not able to correct $ t $ transpositions, then, by
\eqref{eq:disjoint}, there exist distinct codewords $ \myx, \myx' \in \C $
such that $ B(\myx; t) \cap B(\myx'; t) \neq \varnothing $.
By \eqref{eq:distance} and \eqref{eq:mindist}, this implies that
$ \dist_{\min}(\C) \leqslant \dist(\myx, \myx') \leqslant 2t $.

The statement (b) follows from \eqref{eq:zedisjoint} and \eqref{eq:infdist}.
\qed
\end{proof}

The converse of (a) is in general not true.
For example, let $ \C = \{\myx = \mathtt{1 0 1 0 0 0 1 0 1 0}, \linebreak \myy = \mathtt{0 0 1 1 0 0 0 0 1 1}\} \subseteq \Z_2^{10} $.
It is easy to see that the only common string that can be produced by applying
transpositions to $ \myx $ and $ \myy $ is $ \myz = \mathtt{0 1 0 1 0 0 0 1 0 1} $, and
that the only way this can be done is by applying transpositions at locations
$ 1, 3, 7, 9 $ to $ \myx $, and transpositions at locations $ 2, 8 $ to $ \myy $.
This implies that $ \dist_{\min}(\C) = \dist(\myx,\myy) = 4 + 2 = 6 $, and also
that $ B(\myx; 3) \cap B(\myy; 3) = \varnothing $.
In other words, the code $ \C $ corrects $ t = 3 $ transpositions, and yet
$ \dist_{\min}(\C) \leqslant 2t $.

\begin{remark}[Transposition distance]
Another possible definition of ``distance'' between $ \myx $ and $ \myy $ that
may seem natural in this context is the minimum number of transpositions needed
to transform $ \myx $ into $ \myy $ (or, equivalently, $ \myy $ into $ \myx $),
namely $ \dist'(\myx, \myy) \defeq \min\{t \colon \myy \in B(\myx; t) \} $.
However, this version of distance is not appropriate for the problem at hand
because the corresponding analog of Proposition~\ref{thm:eccmetric} does not
hold.
E.g., for $ \myx = \mathtt{1 0 0} $, $ \myy = \mathtt{0 0 1} $, we have
$ \dist'(\myx, \myy) = \infty > 2 $, but $ \{\myx, \myy\} $ is not a code
correcting one transposition because $ \myz = \mathtt{0 1 0} $ can be obtained from
both $ \myx $ and $ \myy $ through one transposition, i.e., $ \myz \in B(\myx; 1) \cap B(\myy; 1) $.
\myqed
\end{remark}

\subsection{A related model -- successively performed transpositions}
\label{sec:succ}

We briefly discuss here another model that is closely related to the one described
in the previous subsection.

If the transmitted string is distorted by $ t $ transpositions in the channel, one
can imagine a scenario where these transpositions are performed in a successive
manner, one after another, and are unrestricted in the sense that they may affect
pairs of symbols that are not necessarily disjoint.
Here is an example with $ \myx, \myy \in \Z_4^{10} $ and $ t = 3 $:
\begin{equation}
\label{eq:xy}
\begin{aligned}
%\label{eq:transp1}
  \myx   \ = \ \  &\mathtt{0 \ 1} \, \aunderbrace[l1r]{\mathtt{1 \ 3}} \, \mathtt{0 \ 0 \ 2 \ 2 \ 2 \ 1}  \\
%\label{eq:transp2}
	        \hookrightarrow   \ &\mathtt{0 \ 1 \ 3} \, \aunderbrace[l1r]{\mathtt{1 \ 0}} \, \mathtt{0 \ 2 \ 2 \ 2 \ 1}  \\
%\label{eq:transp3}
	        \hookrightarrow   \ &\mathtt{0 \ 1 \ 3 \ 0 \ 1 \ 0 \ 2 \ 2} \, \aunderbrace[l1r]{\mathtt{2 \ 1}}  \\
%\label{eq:transp4}
          \hookrightarrow   \ &\mathtt{0 \ 1 \ 3 \ 0 \ 1 \ 0 \ 2 \ 2 \ 1 \ 2}  \ = \  \myy .
\end{aligned}
\end{equation}
In this model, pairs of symbols that are being swapped in two different steps may
overlap, and thus one particular symbol of the transmitted string may be displaced
by up to $ t $ positions away from its original position. %(see \eqref{eq:transp1}--\eqref{eq:transp3}).

Let $ B^{\textnormal{{\tiny (s)}}}(\myx; t) $ denote the set of all strings that can
be obtained by applying $ t $ or fewer successive transpositions to the string $ \myx $.

The definition of distance appropriate for this model is the following:
\begin{align}
\label{eq:dist_succ}
  \dist^{\textnormal{{\tiny (s)}}}(\myx, \myy)
	\defeq  \min\!\big\{ t \colon \myy \in B^{\textnormal{{\tiny (s)}}}(\myx; t) \big\} ,
\end{align}
with the convention $ \min \varnothing = \infty $.
In words, $ \dist^{\textnormal{{\tiny (s)}}}(\myx,\myy) $ is the smallest number of
transpositions, performed in a successive manner, that can transform $ \myx $ into
$ \myy $ (or, equivalently, $ \myy $ into $ \myx $).
If $ \myy $ cannot be obtained from $ \myx $ by using any number of transpositions,
we set $ d^{\textnormal{{\tiny (s)}}}(\myx,\myy) = \infty $.
Unlike $ \dist(\cdot,\cdot) $, $ \dist^{\textnormal{{\tiny (s)}}}(\cdot,\cdot) $ is
a metric%
\footnote{More precisely, it is an extended metric because $ d^{\textnormal{{\tiny (s)}}}(\myx,\myy) = \infty $
if (and only if) the strings $ \myx,\myy $ have different compositions. This metric
was analyzed on the space of permutations over $ \{1, \ldots, n\} $ in, e.g.,
\cite{konstantinova}, but can be defined for arbitrary strings.}.
The set $ B^{\textnormal{{\tiny (s)}}}(\myx; t) $ defined above is a ball of radius
$ t $ around $ \myx $ with respect to this metric.
Defining the minimum distance of a code $ \C \subseteq \Z_q^n $ with respect to
$ \dist^{\textnormal{{\tiny (s)}}} $ in the usual way (see \eqref{eq:mindist}), we
have the following claim.

\begin{proposition}
A code $ \C \subseteq \Z_q^n $ corrects $ t $ successive transpositions if and only if $ \dist^{\textnormal{{\tiny (s)}}}_{\min}(\C) > 2t $.
\end{proposition}
\begin{proof}
The code corrects $t$ successive transpositions if and only if
$B^{\textnormal{{\tiny (s)}}}(\myx;t)$ and
$B^{\textnormal{{\tiny (s)}}}(\myx';t)$ are disjoint for every pair of
distinct codewords $\myx,\myx'$.
If these balls contain a common string $\myz$, the triangle inequality gives
\begin{align*}
d^{\textnormal{{\tiny (s)}}}(\myx,\myx')
&\leqslant d^{\textnormal{{\tiny (s)}}}(\myx,\myz)
       +d^{\textnormal{{\tiny (s)}}}(\myz,\myx')
\leqslant 2t.
\end{align*}
Conversely, suppose that
$d^{\textnormal{{\tiny (s)}}}(\myx,\myx')=m\leqslant2t$.
Choose a shortest sequence of $m$ successive transpositions from $\myx$ to
$\myx'$, and let $\myz$ be the string reached after $\min\{m,t\}$ steps.
Then both $d^{\textnormal{{\tiny (s)}}}(\myx,\myz)\leqslant t$ and
$d^{\textnormal{{\tiny (s)}}}(\myz,\myx')\leqslant t$, so the two balls
intersect. The claim follows.
\qed
\end{proof}

The model with successive transpositions is stronger than the one with simultaneous,
disjoint transpositions in the sense that all error patterns that can be realized
in the latter can also be realized in the former.
Namely, when the pattern of successively performed transpositions is such that they
do not happen to overlap, disjoint transpositions are obtained as a special case.
Therefore, $ B(\myx; t) \subseteq B^{\textnormal{{\tiny (s)}}}(\myx; t) $.
In fact, the stronger statement
$ \bar{B}(\myx;t) \subseteq B^{\textnormal{{\tiny (s)}}}(\myx;t) $
also holds.
Indeed, let $ \myy \in \bar{B}(\myx;t) $.
By \eqref{eq:distance} and \eqref{eq:ball}, there exist nonnegative integers
$ r,s $ with $ r+s \leqslant t $ and a string $ \myz $ such that
$ \myz \in B(\myx;r) \cap B(\myy;s) $.
Thus, $ \myz $ can be obtained from $ \myx $ by at most $ r $ pairwise disjoint
adjacent transpositions, and from $ \myy $ by at most $ s $ pairwise disjoint
adjacent transpositions.
The transpositions within each collection commute and may therefore be performed
successively in any order.
Moreover, since every transposition is its own inverse, the latter collection
can be applied to $ \myz $ to obtain $ \myy $.
Concatenating the two collections therefore gives a sequence of at most
$ r+s\leqslant t $ successive transpositions transforming $ \myx $ into $ \myy $.
Hence $ \myy\in B^{\textnormal{{\tiny (s)}}}(\myx;t) $, proving the claimed
inclusion.

Consequently: (i)~any code construction valid for the successive transposition model
is also valid for the disjoint transposition model, (ii)~any lower bound on the
cardinality of optimal codes for the successive transposition model is also valid
for the disjoint transposition model, and (iii)~any upper bound on the cardinality of
optimal codes in the disjoint transposition model is also valid in the successive
transposition model.
Therefore, the upper bounds given in Theorems \ref{thm:run-upper} and \ref{thm:bounds} below for the model with disjoint transpositions are automatically
valid for the stronger model with successive transpositions as well.%

\subsection{Transpositions vs.\ substitutions}
\label{sec:subvstransp}

For $ \bs{x} \in \Z_q^n $, let $ \bs{\hat{x}} \in \Z_q^n $ be the string defined
by $ \hat{x}_i = \sum_{j=1}^i x_j $, $ i = 1, \ldots, n $, where addition is performed
modulo $ q $.
Equivalently, $ x_i = \hat{x}_i - \hat{x}_{i-1} $, $ i = 1, \ldots, n $, where
$ \hat{x}_0 = 0 $ and subtraction is performed modulo $ q $.
Then it is easy to see that any pattern of at most $ t $ transpositions in $ \bs{x} $
becomes a pattern of at most $ t $ substitutions in $ \bs{\hat{x}} $.
This is true not only for disjoint but for successive transpositions as well.
It follows that, if $ \widehat{\C} \subseteq \Z_q^n $ is a code correcting $ t $
substitutions, i.e., a code having minimum Hamming distance larger than $ 2 t $,
then $ \C = \{\bs{x} \colon \bs{\hat{x}} \in \widehat{\C} \} $ is a code correcting
$ t $ transpositions.
Consequently, any lower bound on the cardinality of optimal codes correcting $ t $
substitutions is automatically a lower bound on the cardinality of optimal codes
correcting $ t $ transpositions.

Since the transposition model is in this sense weaker than the substitution model,
we expect to be able to achieve higher code rates in the former.
This is indeed demonstrated in Section \ref{sec:bounds} and illustrated in Figure \ref{fig:GV}.

\subsection{Prior work}
\label{sec:prior}

The bounded-displacement channel considered here was introduced as the so-called 
$\ell_\infty$-limited permutation channel in \cite{langberg}; the present model
is its radius-one instance.
The works \cite{langberg,chee} studied codes correcting every
admissible displacement pattern and thereby obtained lower bounds on the zero-error
capacity.
The recent works \cite{benshimon+zabokritskiy,benshimon+zabokritskiy2} further
developed the block-concatenation approach from \cite{langberg,chee}.
The first introduced a finite $q$-ary two-stage local criterion that certifies
zero-error correction for concatenations of arbitrary length, and used it to
obtain the binary rate of $\approx\!0.653618$ bits per symbol
\cite{benshimon+zabokritskiy}.
The second introduced the more flexible notion of extinction depth and gave rates of $\approx\!1.074204$ and $\approx\!1.554513$ bits per symbol for
$q=3$ and $q=5$, respectively \cite{benshimon+zabokritskiy2}.
%The rates quoted from \cite{benshimon+zabokritskiy,benshimon+zabokritskiy2} have been converted from base-$q$ normalization to bits per symbol.
Our zero-error construction in Section \ref{sec:ze} uses their local criterion
but extends their symbolic baseline by five new templates in a way that is
valid uniformly in the alphabet size.
It yields rates $\approx\!1.346292$ for $q=4$ and $\approx\!1.571783$ for $q=5$.
(The rates in \cite{benshimon+zabokritskiy,benshimon+zabokritskiy2} are defined and calculated in base $q$, i.e., using $\log_q$; in the present paper we express all rates in bits per symbol and we have therefore used the appropriate conversion to $\log_2$.)
%For growing alphabets ($q\to\infty$), \cite[Theorems~29 and~32]{benshimon+zabokritskiy2} also gives the large-alphabet zero-error bounds
%\begin{align}
% \log_2q-\log_2(1.82560995)+o(1)
% \leqslant R_q^\star(1/2)
% \leqslant \log_2q-\log_2\varphi+o(1),
%\end{align}
%where $\varphi=\frac{1+\sqrt5}{2}$.
%the lower bound is obtained by a run-wise lifting construction and the upper bound by a covering argument.

In the binary case $q = 2$, adjacent transpositions are also known in the literature as bit-shifts or
peak-shifts.
Lower bounds on achievable rates of codes correcting a linear number $ t = \tau n $ of bit-shifts were obtained in
\cite{kolesnik+krachkovsky,goyal}, while constructions and
bounds on codes correcting a fixed number $t = \textrm{const}$ of bit-shifts were given in \cite{kovacevic}.
These works consider the
stronger model with successively performed transpositions/bit-shifts.
Constructions of binary codes correcting a single transposition appear in
some earlier works as well, e.g., in \cite{kuznetsov+hanvinck,levenshtein+hanvinck}.
Thus, previous linear-regime results are confined to the binary successive-transposition model.
To the best of our knowledge, the generalized Gilbert--Varshamov bound and the
run-based upper bound given in this paper are the first nonconstant, $\tau$-dependent
lower and upper bounds developed for the pairwise-disjoint $q$-ary model
throughout the linear regime.
In the $t = \textrm{const}$ regime, asymptotic bounds on the cardinality of optimal
$ q $-ary codes correcting $ t $ transpositions are derived, for any $ q \geqslant 2 $ and
$ t \geqslant 1 $.
Both results hold for the model with pairwise disjoint transpositions.

We mention in this context also the works studying zero-error codes for models
that are not equivalent to the model analyzed here, but are closely related
to it, namely the binary bit-shift channel with an additional assumption that
each $ 1 $ can be shifted by at most $ \ell $ positions \cite{shamai+zehavi,krachkovsky}
and several types of timing channels \cite{kovacevic+popovski,fifo}.

Codes for models that include other types of errors in addition to transpositions,
e.g., deletions and/or substitutions, were studied in
\cite{abdel-ghaffar,chen+hanvinck,damerau,gabrys,gabrys2,khuat,klove,schulman,wang,ye+ge}.

Finally, let us mention that transposition errors are also of interest in another
scenario where the stored data is represented by a \emph{permutation} on
$ \{1, \ldots, n\} $.
Namely, transposition-correcting codes in the space of all permutations occur
naturally in rank modulation schemes for flash memories; see, e.g., \cite{barg+mazumdar}.

\subsection{Notation and conventions}
\label{sec:asymp}

By a \emph{run} in a string $ \myx \in \Z_q^n $ we mean a substring of identical
symbols that is delimited on both sides either by a different symbol or by the end
of the string.
The number of runs in $ \myx $ is denoted by $ \run(\myx) $.
For example, the string $ \myx = \mathtt{0 3 3 3 0 0 1 2 2} \in \Z_4^{9} $ has
$ \run(\myx) = 5 $ runs.

$ H_q(x) = - x \log_q x - (1 - x) \log_q(1 - x) + x \log_q(q-1) $ is the $ q $-ary
entropy function.
As usual, we write simply $ H(x) $ for $ H_2(x) $.

Whenever a rate is displayed in decimal form, it is truncated after six
decimal places.

The binomial coefficient $\binom{a}{b}$ is understood to equal zero when $a < b$.

The minimum of an empty set is taken to be $\infty$ by convention.

We will also need the
following asymptotic notation.
For two nonnegative sequences $ (a_n) $, $ (b_n) $:
\renewcommand\labelitemi{$\vcenter{\hbox{\tiny$\bullet$}}$}
\begin{itemize}[topsep=3pt, itemsep=0pt]
\item
$ a_n \sim b_n $ means $ \lim_{n\to\infty} \frac{a_n}{b_n} = 1 $,
\item
$ a_n \lesssim b_n $ means $ \limsup_{n\to\infty} \frac{a_n}{b_n} \leqslant 1 $,
\item
$ a_n = o(b_n) $ means $ \lim_{n\to\infty} \frac{a_n}{b_n} = 0 $,
\item
$ a_n = {\mathcal O}(b_n) $ means $ \limsup_{n\to\infty} \frac{a_n}{b_n} < \infty $,
\item
$ a_n = \Omega(b_n) $ means $ \liminf_{n\to\infty} \frac{a_n}{b_n} > 0 $, or equivalently $ b_n = {\mathcal O}(a_n) $,
\item
$ a_n = \Theta(b_n) $ means
$ 0 < \liminf_{n\to\infty} \frac{a_n}{b_n} \leqslant \limsup_{n\to\infty} \frac{a_n}{b_n} < \infty $.
\end{itemize}

%--------------------------------------------------

\section{Properties of transpositions and transposition distance}
\label{sec:balls}

In this section we prove several facts about the transposition distance
and the sets $ B(\myx; r) $ and $ \bar{B}(\myx; r) $ that, in addition to being
of interest in their own right, will be needed for the derivation of the bounds
in subsequent sections.

\begin{proposition}
\label{thm:maxdist}
For every $ q \geqslant 2 $, $ n \geqslant 1 $, the maximum distance between
two strings $ \myx, \myy \in \Z_q^n $ such that $ \dist(\myx,\myy) < \infty $ is
$ \max\{\dist(\myx,\myy) \colon \myx,\myy \in \Z_q^n,\, \dist(\myx,\myy) < \infty\} = n - 1 $.
\end{proposition}
\begin{proof}
Suppose a string $ \myz $ can be obtained by applying transpositions at locations
$ k_1, \ldots, k_t $ to the string $ \myx $, and also by applying transpositions
at locations $ \ell_1, \ldots, \ell_s $ to the string $ \myy $.
If $ k_i = \ell_j $, for some $ i \in \{1, \ldots, t\} $, $ j \in \{1, \ldots, s\} $,
then $ (z_{k_i}, z_{k_i+1}) = (x_{k_i+1}, x_{k_i}) = (y_{k_i+1}, y_{k_i}) $, and
hence a common string $ \myz' $ (obtained by applying a transposition at location
$ k_i $ to $ \myz $) can be obtained by applying $ t - 1 $ transpositions (at locations
$ k_1, \ldots, k_{i-1}, k_{i+1}, \ldots k_t $) to $ \myx $ and $ s - 1 $ transpositions
(at locations $ \ell_1, \ldots, \ell_{j-1}, \ell_{j+1}, \ldots \ell_s $) to $ \myy $.
Therefore, when computing $ \dist(\myx,\myy) $, we may, without loss of generality,
assume that the sets of transposition locations $ \{k_1, \ldots, k_t\} $ and
$ \{\ell_1, \ldots, \ell_s\} $ are disjoint.
Since $ k_t, \ell_s \leqslant n-1 $, this implies that $ t + s \leqslant n - 1 $,
and hence that $ \dist(\myx,\myy) \leqslant n - 1 $.

That this upper bound is attained is shown by the following examples (the cases $ n = 1, 2 $ are trivial).
For even $ n \geqslant 4 $, take $ \myx $ to be the alternating string $ \mathtt{1 0 1 0}\squeezeddots $
of length $ n $, and $ \myy $ the concatenation of the string $ \mathtt{0} $, the alternating
string $ \mathtt{0 1 0 1}\squeezeddots $ of length $ n - 2 $, and the string $ \mathtt{1} $.
E.g., for $ n = 6 $, $ \myx = \mathtt{1 0 1 0 1 0} $ and $ \myy = \mathtt{0 0 1 0 1 1} $.
Then the only common string that $ \myx $ and $ \myy $ can produce is the alternating
string $ \myz = \mathtt{0 1 0 1}\squeezeddots $, obtained by applying $ n/2 $ transpositions at
locations $ 1, 3, \ldots, n-1 $ to $ \myx $, and $ n/2 - 1 $ transpositions at locations
$ 2, 4, \ldots, n-2 $ to $ \myy $.
To see this, read the strings from left to right.
The first two symbols of $\myy$ are $\mathtt{0}$, so every common output
starts with $\mathtt{0}$; this forces a transposition at location $1$ in
$\myx$. Its second output symbol is then $\mathtt{1}$, which forces a
transposition at location $2$ in $\myy$.  Continuing alternately forces all
the odd locations in $\myx$ and all the even locations in $\myy$, and hence
forces the alternating common output.  The total number of transpositions is
$n-1$.
Similarly, for odd $ n \geqslant 3 $, take $ \myx $ to be the concatenation of the alternating
string $ \mathtt{1 0 1 0}\squeezeddots $ of length $ n - 1 $ and the string $ \mathtt{0} $, and $ \myy $
the concatenation of the string $ \mathtt{0} $ and the alternating string $ \mathtt{0 1 0 1}\squeezeddots $.
E.g., for $ n = 5 $, $ \myx = \mathtt{1 0 1 0 0} $ and $ \myy = \mathtt{0 0 1 0 1} $.
\qed
\end{proof}

%--------------------------------------------------

Of particular interest for the present paper are the cardinalities of the sets
$ B(\myx; r) $ and $ \bar{B}(\myx; r) $.
Obtaining the exact characterization of these quantities appears to be difficult,
but one can derive good bounds.

First note that
\begin{align}
\label{eq:B1}
  |B(\myx; 1)| = |\Beq(\myx; 1)| + 1 = \run(\myx) .
\end{align}
This is because only the transpositions occurring at the boundaries of consecutive
runs in $ \myx $ will actually alter $ \myx $, and hence there are $ \run(\myx) - 1 $
strings different from $ \myx $ that can be obtained in this way.
Accounting for $ \myx $ itself, \eqref{eq:B1} follows.
For general $ r $, we have the bounds on $ |\Beq(\myx; r)| $ stated in Proposition
\ref{thm:cardB}.
Bounds on $ |B(\myx; r)| $ can then be obtained from
\begin{align}
\label{eq:BBeq}
  |B(\myx; r)|  =  \sum_{r'=0}^{r} |\Beq(\myx; r')| ,
\end{align}
which follows from \eqref{eq:BBequnion}.

Let $M\subseteq\{1,\ldots,n-1\}$ be an admissible set of
transposition locations, meaning that no two locations are consecutive (see Section~\ref{sec:model}).
Write $\sigma_M\myx$ for the resulting string.
We call $M$ \emph{effective} for $\myx$ if $x_i\neq x_{i+1}$ for every
$i\in M$.  Every word obtainable from $\myx$ has a unique effective set.
Indeed, ineffective transpositions may first be deleted.  If two distinct
effective sets $M,M'$ produced the same word, let $i$ be the smallest
location in their symmetric difference and suppose, without loss of
generality, that $i\in M\setminus M'$.  The spacing condition and the
minimality of $i$ imply that $M'$ contains neither $i-1$ (when $i>1$) nor
$i$.  Position $i$ of $\sigma_M\myx$ therefore contains $x_{i+1}$, whereas
position $i$ of $\sigma_{M'}\myx$ contains $x_i$, contradicting the
effectiveness of $M$.  Consequently, if $M$ is effective and $|M|=r$, then
$\sigma_M\myx\in\Beq(\myx;r)$.

\begin{proposition}
\label{thm:cardB}
For every $ q \geqslant 2 $, $ n \geqslant 1 $, $ \myx \in \Z_q^n $, $ r \geqslant 1 $,
\begin{align}
\label{eq:cardB}
  \binom{\run(\myx)-r}{r}
  \;\leqslant\;
  |\Beq(\myx; r)|
  \;\leqslant\;
  \min\left\{\binom{\run(\myx)-1}{r},\binom{n-r}{r}\right\} .
\end{align}
\end{proposition}
\begin{proof}
Put $m=\run(\myx)$.  Form a graph $G_{\myx}$ whose $m-1$ vertices are the
boundaries between consecutive runs of $\myx$, in their natural order, and
join two consecutive boundaries exactly when the run between them has length
one.  Thus $G_{\myx}$ is a subgraph of the path $P_{m-1}$.  A set of run
boundaries specifies pairwise disjoint transpositions if and only if it is an
independent set of $G_{\myx}$.  All these transpositions are effective, and
the effective-set uniqueness established above shows that distinct
independent sets give distinct elements of $\Beq(\myx;r)$.  Conversely,
every element of $\Beq(\myx;r)$ has a unique effective set consisting of
$r$ run boundaries.  Therefore,
writing $i_r(G)$ for the number of independent $r$-sets of a graph $G$, we have
\begin{align}
\label{eq:Beq-independent}
  |\Beq(\myx;r)|=i_r(G_{\myx}).
\end{align}
Since $G_{\myx}\subseteq P_{m-1}$, every independent $r$-set of the path is
also independent in $G_{\myx}$.  The standard count of nonconsecutive
$r$-subsets gives
\begin{align}
\label{eq:path-independent}
  i_r(P_{m-1})=\binom{m-r}{r},
\end{align}
which proves the lower bound.

On the other hand, an independent $r$-set is
an $r$-subset of the $m-1$ run boundaries and also an $r$-subset of
$\{1,\ldots,n-1\}$ containing no consecutive locations.  There are
$\binom{m-1}{r}$ sets of the former kind and $\binom{n-r}{r}$ of the latter,
which proves the upper bound.
\qed
\end{proof}

\begin{proposition}
\label{thm:Bupper}
For every $ q \geqslant 2 $, $ n \geqslant 1 $, $ \myx \in \Z_q^n $, $ r \geqslant 1 $,
\begin{align}
\label{eq:Bupper}
  |B(\myx; r)|
  \;\leqslant\; |\bar{B}(\myx; r)|
  \;\leqslant\;
  \sum_{u=\lceil r/3 \rceil}^{r}
  \binom{2\min\{u,\lfloor n/2\rfloor\}}{r-u}
  | B(\myx; u) | .
\end{align}
\end{proposition}
\begin{proof}
The left-hand inequality follows from
$B(\myx;r)\subseteq\bar{B}(\myx;r)$.

Take any $\myw\in\bar{B}(\myx;r)$.
There are sets of transpositions $M,N$ such that
\begin{align}
  \myw=\sigma_N\sigma_M\myx,
  \qquad |M|+|N|\leqslant r .
\end{align}
This representation can be normalized without increasing $|M|+|N|$.
First, any transposition in $M \cap N$ can be deleted from both sets, because it commutes
with every other transposition and the two copies cancel.
Further, if an element $j\in N$ determines a transposition that is disjoint from every transposition determined by $M$, i.e., $\min\{|j - i| \colon i \in M\} \geqslant 2$, it can be moved into $M$.
Finally, any ineffective transposition in $M$ (that exchanges equal symbols of $\myx$) can be deleted.
Repeating these operations leaves $M$ effective and makes every transposition in $N$
partially overlap with a transposition in $M$, i.e., $\forall j \in N \ \exists i \in M \ \text{s.t.} \ |j - i| = 1$.

Write $k=|M|$ and $v=|N|$.
There are at most $2k$ locations partially overlapping the transpositions from $M$.
For a fixed intermediate string $\sigma_M\myx\in\Beq(\myx;k)$, whose effective
set of transpositions is unique, the set $N$ can therefore be chosen in at most
$\binom{2k}{v}$ ways.
Consequently,
\begin{align}
  |\bar{B}(\myx;r)|
  &\leqslant
  \sum_{k=0}^{\lfloor n/2\rfloor}|\Beq(\myx;k)|
  \sum_{\substack{0\leqslant v\leqslant 2k\\ k+v\leqslant r}}
  \binom{2k}{v}                                                   \notag\\
  &=
  \sum_{u=0}^{r}\sum_{k=0}^{\min\{u,\lfloor n/2\rfloor\}}
  |\Beq(\myx;k)|\binom{2k}{r-u},
  \label{eq:Bupper-intermediate}
\end{align}
where $u=r-v$.
For fixed $u$, set $s=\min\{u,\lfloor n/2\rfloor\}$.
Since the binomial coefficient is nondecreasing in its upper argument,
\begin{align}
\nonumber
  \sum_{k=0}^{s}|\Beq(\myx;k)|\binom{2k}{r-u}
  &\leqslant
  \binom{2s}{r-u}\sum_{k=0}^{s}|\Beq(\myx;k)|  \\
  &=\binom{2s}{r-u}|B(\myx;s)| .
\end{align}
If $u<r/3$, then $r-u>2u\geqslant2k$ for every $k\leqslant u$, so the
corresponding terms in \eqref{eq:Bupper-intermediate} vanish.
This proves the right-hand inequality in \eqref{eq:Bupper}.
\qed
\end{proof}

We next define and analyze the average cardinalities of the sets $ \Beq(\myx; r) $,
$ B(\myx; r) $, and $ \bar{B}(\myx; r) $ over the whole space $ \Z_q^n $.
Let
\begin{subequations}
\label{eq:totalballs}
\begin{align}
  \Teq_q(n; r)  &\defeq  \sum_{\myx \in \Z_q^n} |\Beq(\myx; r)| ,  \\
  T_q(n; r)  &\defeq  \sum_{\myx \in \Z_q^n} |B(\myx; r)| ,  \\
  \bar{T}_q(n; r)  &\defeq  \sum_{\myx \in \Z_q^n} |\bar{B}(\myx; r)| ,
\end{align}
\end{subequations}
so that the average value of $ |\Beq(\myx; r)| $ is $ \frac{1}{q^n} \Teq_q(n; r) $,
and similarly for the others.
It follows from \eqref{eq:BBeq}, \eqref{eq:Bupper}, and \eqref{eq:totalballs} that
\begin{align}
\label{eq:PT}
  T_q(n; r) = \sum_{r'=0}^{r} \Teq_q(n; r')
%	\;\leqslant\; (r+1) \max_{0\leqslant r'\leqslant r} T^{\sml{=}}_q(n; r)
\end{align}
and
\begin{align}
\label{eq:TTbar}
  {T}_q(n; r)
		\leqslant  \bar{T}_q(n; r)
		\leqslant
  \sum_{u=\lceil r/3 \rceil}^{r}
  \binom{2\min\{u,\lfloor n/2\rfloor\}}{r-u}
  T_q(n;u) .
\end{align}

\begin{proposition}
\label{thm:Teq}
For every $ q \geqslant 2 $, $ n \geqslant 1 $, $ r \geqslant 0 $,
\begin{align}
\label{eq:Teq}
  \Teq_q(n;r) = \binom{n-r}{r} q^{n-r} (q-1)^r .
\end{align}
\end{proposition}
\begin{proof}
By the effective-set uniqueness established in the paragraph before
Proposition~\ref{thm:cardB},
$\Teq_q(n;r)$ counts pairs $(\myx,M)$ in which $M$ is an effective set of
$r$ transposition locations for $\myx$.  There are $\binom{n-r}{r}$ sets of
$r$ pairwise nonconsecutive locations in $\{1,\ldots,n-1\}$.  For each fixed
set, its $r$ pairs of coordinates are disjoint; there are $q(q-1)$ choices
for the two unequal symbols in every pair and $q$ choices at each of the
remaining $n-2r$ coordinates.  Therefore
\begin{align}
\label{eq:Teq-direct}
  \Teq_q(n;r)
  =\binom{n-r}{r}[q(q-1)]^r q^{n-2r}
  =\binom{n-r}{r}q^{n-r}(q-1)^r,
\end{align}
as claimed.
\qed
\end{proof}

\begin{remark}[Generating-function derivation]
\textnormal{
Alternatively, one can determine the expression for $\Teq_q(n;r)$ via its generating function.
Conditioning on the last two symbols of the input, we get:
\begin{itemize}[topsep=3pt, itemsep=0pt]
\item
If the last two symbols of $ \myx \in \Z_q^n $ are different, i.e., $ \bs{x} = \bs{x}' y y' $
with $ \myx' \in \Z_q^{n-2} $, $ y, y' \in \Z_q $, $ y \neq y' $, then
\begin{align}
  |\Beq(\bs{x}; r)| = |\Beq(\bs{x}' y; r)| + |\Beq(\bs{x}'; r-1)| .
\end{align}
The first term corresponds to the case when all the transpositions occur in the
substring $ \bs{x}' y $, and the second term to the case when $ r-1 $ transpositions
occur in the substring $ \bs{x}' $ and one transposition occurs in $ y y' $.
\item
If the last two symbols of $ \myx \in \Z_q^n $ are the same, i.e., $ \bs{x} = \bs{x}'yy $,
then
\begin{align}
  |\Beq(\bs{x}; r)| = |\Beq(\bs{x}' y; r)| .
\end{align}
\end{itemize}
The following
recurrence is then obtained
\begin{align}
\label{eq:Trec}
  \Teq_q(n;r)
  =q\Teq_q(n-1;r)+q(q-1)\Teq_q(n-2;r-1),
\end{align}
with the natural initial conditions
$\Teq_q(0;0) = 1$, $\Teq_q(0;r>0) = 0$, $\Teq_q(1;0) = q$, $\Teq_q(1;r>0) = 0$,
from which one finds the bivariate generating function of this sequence in the form
\begin{align}
\nonumber
  F^{\textnormal{e}}_q(z,w)
	\defeq \sum_{n,r\geqslant 0} \Teq_q(n; r) z^{n} w^r
  &=  \frac{1}{1 - qz - q(q-1)z^2 w}  \\
\label{eq:F2}
  &=  \frac{1}{1 - qz} \sum_{r=0}^\infty \left( \frac{q(q-1)z^2}{1-qz} \right)^r w^r  .
\end{align}
Expanding the right-hand side first in $w$ and then in $z$ gives
\eqref{eq:Teq} directly, namely $\Teq_q(n;r) = [z^n w^r] F^{\textnormal{e}}_q(z,w)$.
\myqed}
\end{remark}

We end this section with a lemma we shall need in the sequel. It gives the exponential growth rate of the number of strings with
relatively few runs.

\begin{lemma}
\label{thm:rho}
For fixed $ \mu \in (0, 1] $,
\begin{align}
\label{eq:Hq}
  \lim_{n \to \infty} \frac{1}{n} \log_q \!\left| \{\myx \in \Z_q^n \colon \run(\myx) \leqslant \mu n \} \right| =
		\begin{cases}
		 H_q(\mu) , & 0 < \mu \leqslant 1 - q^{-1} \\
		 1 ,        & 1 - q^{-1} < \mu \leqslant 1
		\end{cases},
\end{align}
where $ H_q(\cdot) $ is the $ q $-ary entropy function.
\end{lemma}
\begin{proof}
Let $ S_q(n,m) $ denote the number of strings from $ \Z_q^n $ having exactly
$ m $ runs of identical symbols, so that
\begin{align}
\label{eq:Sqrn}
  \left| \{\myx \in \Z_q^n \colon \run(\myx) \leqslant \mu n \} \right|
  =  \sum_{m=1}^{\lfloor \mu n \rfloor} S_q(n,m) .
\end{align}
By a direct counting argument we find that
\begin{align}
\label{eq:Sqexact}
  S_q(n,m) = \binom{n-1}{m-1} q (q-1)^{m-1}
\end{align}
and hence, by Stirling's approximation,
\begin{align}
\label{eq:Sqexponent}
  \lim_{n\to\infty} \frac{1}{n} \log_q S_q(n,\lfloor \mu n \rfloor)  =  H_q(\mu) .
\end{align}
Since the quantity we are interested in, expressed in \eqref{eq:Sqrn}, can be
sandwiched between
\begin{align}
\label{eq:Sqmax-sandwich}
  \max_{m \leqslant \lfloor \mu n \rfloor} S_q(n,m)
		\leqslant
  \sum_{m=1}^{\lfloor \mu n \rfloor} S_q(n,m)
		\leqslant
  \mu n \cdot \max_{m \leqslant \lfloor \mu n \rfloor} S_q(n,m) ,
\end{align}
its exponential growth rate (to base $ q $) is
\begin{align}
\label{eq:Sqball-exponent}
  \lim_{n\to\infty} \frac{1}{n} \log_q \sum_{m=1}^{\lfloor \mu n \rfloor} S_q(n,m)
  =  \max_{0 \leqslant \mu' \leqslant \mu} H_q(\mu') .
\end{align}
That this is equal to the right-hand side of \eqref{eq:Hq} follows from the fact
that $ H_q(\mu) $ is a concave function of $ \mu \in [0,1] $ maximized at
$ \mu = 1 - q^{-1} $, the maximum being $ H_q(1 - q^{-1}) = 1 $.
\qed
\end{proof}

%--------------------------------------------------

\section{Bounds on codes correcting $ t \!=\! \tau n $ transpositions}
\label{sec:bounds}

Let $ R_q^\star(\tau) $ be the largest asymptotic rate of codes in $ \Z_q^n $
correcting a fraction of $ \tau \in [0, \frac{1}{2}] $ transpositions, expressed
in bits per symbol,
\begin{align}
\label{eq:Rstar}
  {R}_q^\star(\tau) \defeq \limsup_{n\to\infty} \frac{1}{n} \log_2 {M}_q^\star(n; \lfloor \tau n \rfloor) .
\end{align}
In this section we derive two lower bounds and one upper bound on the function
$ R_q^\star(\tau) $.  The first lower bound is a generalized
Gilbert--Varshamov bound.  The second is a lower bound on the zero-error
capacity of our model, i.e., a lower bound on $ R_q^\star(1/2) $, and follows from a
construction of codes correcting all possible patterns of adjacent
transpositions.  The upper bound is obtained by splitting a code according to
the number of runs in its codewords and then applying a packing argument to
the high-run part.

%--------------------------------------------------

\subsection{Generalized Gilbert--Varshamov bound}
\label{sec:GV}

The standard Gilbert--Varshamov bound states that the cardinality of an optimal
code of minimum distance $ \delta $ is lower-bounded by the ratio of the cardinality
of the code space and the cardinality of a ball of radius $ \delta - 1 $ \cite{gilbert,varshamov}.
The bound was later generalized \cite{kolesnik+krachkovsky2,gu+fuja} to include
nonuniform spaces in which the cardinality of a ball of given radius depends on
its center.
Namely, it was proved in \cite{gu+fuja} that, in such cases, the lower bound
still holds if one puts in the denominator the \emph{average} cardinality of a
ball of radius $ \delta - 1 $.

For $ q \geqslant 2 $ and $ \rho \in [0, \frac{1}{2}] $, define
\begin{equation}
\label{eq:alpha}
  \beta^{\textnormal{e}}_q(\rho)  \defeq
	(1-\rho) H\Big(\frac{\rho}{1-\rho}\Big) + (1-\rho)\log_2(q) + \rho\log_2(q-1) ,
\end{equation}
where $ H(\cdot) $ is the binary entropy function.
It is easy to check, by computing its derivatives, that the function
$ \beta^{\textnormal{e}}_q(\rho) $ is concave and maximized for
\begin{align}
   \rho^* = \frac{1}{2} \left( 1 - \sqrt{\frac{q}{5q-4}} \right) .
\end{align}
Further, for $ q \geqslant 2 $ and $ \rho \in [0, \frac{1}{2}] $, define
\begin{equation}
\label{eq:beta}
  \beta_q(\rho)  \defeq  \max_{0\leqslant\rho'\leqslant\rho} \beta^{\textnormal{e}}_q(\rho') =
	\begin{cases}
	  \beta^{\textnormal{e}}_q(\rho) ,   & \rho \leqslant \rho^*  \\
		\beta^{\textnormal{e}}_q(\rho^*) , & \rho \geqslant \rho^*
	\end{cases}.
\end{equation}

\begin{theorem}
\label{thm:GV}
For any $ q \geqslant 2 $ and $ \tau \in [0, \frac{1}{2}] $,
\begin{align}
\label{eq:GV4}
  {R}^\star_q(\tau)  \;\geqslant\;  2 \log_2(q) - \max_{\frac{2\tau}{3}\leqslant\lambda\leqslant\min\{2\tau,\frac{1}{2}\}}  \left[ 2 \lambda H\Big(\frac{2\tau - \lambda}{2\lambda}\Big) + \beta_q(\lambda) \right] ,
\end{align}
where the function $ \beta_q(\cdot) $ is defined in \eqref{eq:beta}.
At $\tau=0$, the right-hand side of \eqref{eq:GV4} is by continuity taken to be $\log_2 q$. 
%the maximand in \eqref{eq:GV4} is understood by continuity at $\lambda=0$, so the right-hand side equals $\log_2(q)$.
\end{theorem}
\begin{proof}
As noted in Section \ref{sec:model}, the distance function $d(\cdot,\cdot)$ is
not a metric. Nevertheless, the required finite-length bound follows directly
from a graph-theoretic form of the Gilbert--Varshamov argument
\cite{tolhuizen}. For fixed $n$ and $t$, let $G_{n,t}$ be the simple graph with
vertex set $\Z_q^n$ in which two distinct strings $\myx,\myy$ are adjacent if
and only if $d(\myx,\myy)\leqslant 2t$. Every independent set in $G_{n,t}$
has minimum distance larger than $2t$ and hence, by Proposition
\ref{thm:eccmetric}, is a code correcting $t$ transpositions. Therefore,
$M_q^\star(n;t)\geqslant\alpha(G_{n,t})$.

Tur\'an's bound on the independence number gives
\begin{align}
  \alpha(G_{n,t})\geqslant
  \frac{|V(G_{n,t})|^2}{|V(G_{n,t})|+2|E(G_{n,t})|}.
\end{align}
On the other hand, by the definition of $G_{n,t}$,
\begin{align}
  \deg_{G_{n,t}}(\myx)+1=|\bar B(\myx;2t)|,
\end{align}
and consequently
\begin{align}
  |V(G_{n,t})|+2|E(G_{n,t})|
  =\sum_{\myx\in\Z_q^n}|\bar B(\myx;2t)|
  =\bar T_q(n;2t).
\end{align}
Since $|V(G_{n,t})|=q^n$, we obtain
\begin{align}
\label{eq:GV1}
  {M}_q^\star(n; t)
  \geqslant \frac{q^{2n}}{\bar{T}_q(n;2t)} ,
\end{align}

If we denote the exponential growth rate (to base $ 2 $) of $ \bar{T}_q(n; \lfloor \rho n \rfloor) $
by
\begin{align}
  \bar{\beta}_q(\rho) \defeq \lim_{n\to\infty} \frac{1}{n} \log_2 \bar{T}_q(n; \lfloor \rho n \rfloor) ,
\end{align}
then \eqref{eq:GV1} implies that
\begin{align}
\label{eq:GV3}
  {R}^\star_q(\tau)  \geqslant  2 \log_2(q) - \bar{\beta}_q(2\tau) .
\end{align}
%The upper-bound calculation below applies equally to any integer sequence $r_n$ satisfying $r_n/n\to\rho$. Thus the bounded discrepancy between $2\lfloor\tau n\rfloor$ and $\lfloor2\tau n\rfloor$ does not affect the exponent in \eqref{eq:GV3}.
Therefore, an upper bound on the function $ \bar{\beta}_q(\cdot) $ would imply
a lower bound on the achievable asymptotic rates of transposition-correcting codes.

From Proposition \ref{thm:Teq} and the fact that
$ \lim_{n\to\infty} \frac{1}{n} \log_2\!\binom{\alpha n}{\gamma n} = \alpha H(\frac{\gamma}{\alpha}) $
one can obtain the exponential growth rate of $ \Teq_q(n; \lfloor \rho n \rfloor) $,
namely
\begin{align}
\label{eq:alphadef}
  \lim_{n\to\infty} \frac{1}{n} \log_2 \Teq_q(n; \lfloor \rho n \rfloor) = \beta^{\textnormal{e}}_q(\rho) ,
\end{align}
where $ \beta^{\textnormal{e}}_q(\rho) $ is given by \eqref{eq:alpha}.
Note that, since the maximum number of disjoint transpositions is $ \lfloor n/2 \rfloor $,
the domain of this function is $ [0, \frac{1}{2}] $.
Then from \eqref{eq:PT} one obtains the exponential growth rate of
$ T_q(n; \lfloor \rho n \rfloor) $, namely
\begin{align}
  \lim_{n\to\infty} \frac{1}{n} \log_2 T_q(n; \lfloor \rho n \rfloor) = \beta_q(\rho) ,
\end{align}
where $ \beta_q(\rho) $ is given by \eqref{eq:beta}.

%Finally, put $m=\lfloor n/2\rfloor$ and $r_0=\min\{r,2m\}$.  Each of the two transposition layers contains at most $m$ transpositions, and therefore
%\begin{align}
%\label{eq:radius-saturation}
%  \bar T_q(n;r)=\bar T_q(n;r_0).
%\end{align}
Proposition~\ref{thm:Bupper} implies that, for any $r \leqslant 2\lfloor n/2\rfloor$,
\begin{align}
\label{eq:Tupper}
  \bar{T}_q(n; r)
		\leqslant  \Big( \frac{2r}{3} + 1 \Big)
  \max_{\lceil r/3\rceil \leqslant u \leqslant \min\{r, \lfloor n/2\rfloor\}}
  \binom{2u}{r-u}T_q(n;u) .
\end{align}
To justify the restricted maximum in \eqref{eq:Tupper}, consider a summand in
\eqref{eq:TTbar} with $u>\lfloor n/2\rfloor$.  Since $r\leqslant 2\lfloor n/2\rfloor$, we have
$0\leqslant r-u<r-\lfloor n/2\rfloor\leqslant \lfloor n/2\rfloor$, and hence
$\binom{2\lfloor n/2\rfloor}{r-u}\leqslant\binom{2\lfloor n/2\rfloor}{r-\lfloor n/2\rfloor}$.
Thus every omitted summand is
dominated by the summand with $u=\lfloor n/2\rfloor$ (both have the factor $T_q(n;\lfloor n/2\rfloor)$).
The total number of summands is at most
$2r/3+1$, proving \eqref{eq:Tupper}.
%This also handles the endpoint $\rho=1$: when $n$ is odd, radii $n$ and $n-1$ give the same set by \eqref{eq:radius-saturation}.
It follows that, for any $ q \geqslant 2 $ and $ \rho \in [0, 1] $,
\begin{equation}
\label{eq:betabar}
  \bar{\beta}_q(\rho)  \leqslant
  \max_{\frac{\rho}{3} \leqslant \lambda \leqslant \min\{\rho,\frac{1}{2}\}}  \left[ 2 \lambda H\Big(\frac{\rho - \lambda}{2\lambda}\Big) + \beta_q(\lambda) \right] .
\end{equation}
The theorem follows from \eqref{eq:GV3} and \eqref{eq:betabar}.
\qed
\end{proof}

The bound from Theorem \ref{thm:GV} is plotted in Figure \ref{fig:GV} for an
alphabet of size $ q = 4 $.
For comparison, we also plot the bound
\begin{align}
\label{eq:GVsub}
  {R}^\star_q(\tau)  \geqslant
	\begin{cases}
	  \log_2(q) - 2\tau\log_2(q-1) - H(2\tau) , &  0 \leqslant \tau \leqslant \frac{q-1}{2q}  \\
		0 , &  \tau > \frac{q-1}{2q}
	\end{cases} .
\end{align}
The expression on the right-hand side of \eqref{eq:GVsub} is the familiar
Gilbert--Varshamov bound for codes correcting $ \tau n $ \emph{substitutions},
i.e., codes of minimum Hamming distance $ \sim\!2\tau n $ \cite[Section 2.10.6]{huffman+pless}.
Recall from Section \ref{sec:subvstransp} that any code correcting $ t $
substitutions can be used to construct a code of the same cardinality correcting
$ t $ transpositions.

\begin{remark}[Lower bound for the binary case]
\label{rem:binarylin}
\textnormal{
For $ q = 2 $, a lower bound on $ {R}^\star_2(\tau) $ better than \eqref{eq:GV4}
is known \cite[Section IV.C]{goyal}.
This is due to the fact that binary codes correcting transposition (a.k.a.\ bit-shift)
errors can be interpreted as packings in the Manhattan metric \cite{kovacevic},
which in particular leads to a finer estimate of $ \bar{T}_2(n;r) $.
Unfortunately, the mentioned geometric interpretation fails when $ q > 2 $ and
thus the corresponding analysis and the constructions via dense packings cannot
be generalized to larger alphabets in any obvious way.
\myqed}
\end{remark}

%------------------------------------------------

\subsection{Codes correcting all patterns of transpositions and the zero-error bound}
\label{sec:ze}

It turns out that exponentially large codes of minimum distance $\infty$ exist
in the model studied in this paper.
The zero-error capacity of the model is the largest asymptotic rate of codes
correcting every admissible pattern of pairwise disjoint adjacent
transpositions.
By \eqref{eq:infdist}, Proposition~\ref{thm:eccmetric}(b), and
\eqref{eq:Rstar}, this rate is precisely $R_q^\star(1/2)$.

For $q=2$, the lower bound
$R_2^\star(1/2)\geqslant0.587430\squeezeddots$ was obtained in \cite[Corollary 20]{langberg} and improved to
$ R_2^\star(1/2) \geqslant 0.642805\squeezeddots $
in \cite[Theorem 3]{chee}.
Both results were obtained by a variable-length block-concatenation construction. 
The general-alphabet construction of \cite{chee} gives
\begin{align}
\label{eq:Rinfty}
  R_q^\star(1/2) \geqslant
  \log_2\!\sqrt{\lfloor q/2\rfloor\!\cdot\!\lceil q/2\rceil}.
\end{align}
More recently, the authors of \cite{benshimon+zabokritskiy} developed a finite
$q$-ary criterion for certifying such block sets and improved the binary lower bound
to $R_2^\star(1/2)\geqslant 0.653618\squeezeddots$, very close to the upper bound $2/3$.
Their subsequent paper \cite{benshimon+zabokritskiy2} gives rates
$R_3^\star(1/2)\geqslant 1.074204\squeezeddots$ for $q=3$ and $R_5^\star(1/2)\geqslant 1.554513\squeezeddots$ for $q=5$.
(The $q=4$ improvement reported there concerns error detection rather than
error correction.)
Restricting the eleven-template baseline used in \cite{benshimon+zabokritskiy2} to four
symbols gives the rate $R_4^\star(1/2) \geqslant 1.324774\squeezeddots$ for $q=4$.

Note that any lower bound on the zero-error capacity is automatically a lower bound on
$R_q^\star(\tau)$ for every $\tau\in[0,\frac12]$.
Thus, writing $R_q^{\textsc{gv}}(\tau)$ for the right-hand side of
\eqref{eq:GV4}, we have
\begin{align}
\label{eq:joint}
  R_q^\star(\tau)
  \geqslant
  \max\left\{R_q^{\textsc{gv}}(\tau),  R_q^\star(1/2)\right\}.
\end{align}

\subsubsection*{The finite certification criterion.}
We first recall the part of the two-stage criterion from
\cite[Definition 11 and Theorem 3]{benshimon+zabokritskiy} that we shall use.
For a string $\myw\in\Z_q^\ell$, put
\begin{align}
  \mathsf B(\myw)&\defeq B(\myw;\lfloor\ell/2\rfloor),\\
  \mathsf T(\myw)&\defeq
  \big\{\operatorname{pref}_{\ell-1}(\myz) \colon \myz\in\mathsf B(\myw)\big\},
\end{align}
where $\operatorname{pref}_j(\myw)$ denotes the length-$j$ prefix of $\myw$.
For equal-length strings $\myu,\myv$, define
\begin{align}
  D_{\min}(\myu,\myv)
  \defeq
  \min\big\{d_{\mathrm H}(\myu',\myv') \colon
  \myu'\in\mathsf B(\myu),\ \myv'\in\mathsf B(\myv)\big\}.
\end{align}
For a finite set of nonempty blocks $\mathcal P$, let $\mathcal P^*$ be
its free concatenation language and let
$\operatorname{Pref}_d(\mathcal P^*)$ be the set of length-$d$ prefixes
of words in that language.

The criterion states that every
$\mathcal P^*\cap\Z_q^n$ is a zero-error code if the following conditions
hold:
\begin{itemize}[topsep=3pt,itemsep=2pt]
\item[(i)]
$\mathsf T(\myu)\cap\mathsf T(\myv)=\varnothing$ for every two distinct
equal-length blocks $\myu,\myv\in\mathcal P$;
\item[(ii)]
whenever $\myx,\myy\in\mathcal P$ and $|\myx|<|\myy|$, writing
$b=|\myy|-|\myx|$ and $\myy_0=\operatorname{pref}_{|\myx|}(\myy)$, one has
$D_{\min}(\myx,\myy_0)\neq0$, and, if
$D_{\min}(\myx,\myy_0)=1$, then
\begin{align}
  \mathsf T(\myx\bs{\rho})\cap\mathsf T(\myy)=\varnothing
  \qquad
  \forall\bs{\rho}\in
  \operatorname{Pref}_b(\mathcal P^*).
\end{align}
\end{itemize}
These conditions also imply that $\mathcal P$ is prefix-free
\cite[Lemma 4]{benshimon+zabokritskiy}.
The usefulness of the criterion is that it replaces a statement about
arbitrarily long concatenations by a finite list of local tests.

\subsubsection*{An alphabet-uniform template family.}
A formal template is a word over the letters $\mathtt{a,b,c,d}$.
For such a template $w$, let $\langle w\rangle_q$ denote the set of words
obtained by assigning distinct symbols of $\Z_q$ to distinct letters occurring
in $w$.
Let $\mathcal T$ denote the following set of sixteen templates, grouped
according to their lengths:
\begin{equation}
\label{eq:uniform-templates}
\begin{array}{c|l}
3 & \mathtt{aaa}\\
4 & \mathtt{abbb}\\
6 & \mathtt{aabbbb},\ \mathtt{aabbcc},\ \mathtt{abaccc},\
    \mathtt{abcbdd}\\
7 & \mathtt{abcccaa}\\
8 & \mathtt{aabbbcaa},\ \mathtt{abacbbbb},\ \mathtt{abccaabb},\
    \mathtt{abccadbb},\\
  & \mathtt{aabccdaa},\ \mathtt{aabbcbdd},\ \mathtt{abccdaaa},\
    \mathtt{abcadbaa},\ \mathtt{abcdddbb}.
\end{array}
\end{equation}
The first eleven are the symbol-symmetric baseline from
\cite{benshimon+zabokritskiy2}; the final five are new.
Let
\begin{align}
\label{eq:Pq}
  \mathcal P_q
  \defeq
  \bigcup_{w\in\mathcal T}\langle w\rangle_q,
  \qquad
  \D_q(n)\defeq\mathcal P_q^*\cap\Z_q^n .
\end{align}
In the following statement we use the falling-factorial notation
$(q)_k=q(q-1)\cdots(q-k+1)$, with $(q)_k=0$ for $k>q$.

\begin{theorem}
\label{thm:Rinfty}
For every $q\geqslant2$, the set $\D_q(n)$, whenever nonempty, is a code correcting an
arbitrary number of pairwise disjoint adjacent transpositions.
Consequently,
\begin{align}
\label{eq:Rinftynew}
  R_q^\star(1/2)\geqslant\log_2(\lambda_q),
\end{align}
where $\lambda_q$ is the unique positive solution of
\begin{align}
\label{eq:lambdaq}
x^8 - q x^{5} -(q)_2 x^{4}
- \big((q)_2+2(q)_3+(q)_4\big)x^{2} -(q)_3 x
- \big(3(q)_3+6(q)_4\big) = 0.
\end{align}
\end{theorem}
\begin{proof}
We verify the finite criterion stated above.
The channel action and every test in the criterion depend only on which symbol
positions are equal.  We therefore work with joint equality patterns rather
than with a fixed alphabet.  First, the distinct symbols in the first block
(the shorter one in unequal-length tests) are assigned canonical labels.
For the other block, we enumerate every way
in which its formal letters can be assigned injectively either to distinct
labels already present or to fresh labels; fresh labels are introduced in
canonical order.  In the second-stage tests, we likewise enumerate the
internal equality pattern of every legal continuation prefix, every possible
identification of its symbols with labels already present in the two blocks,
and as many additional fresh labels as are needed.

Every individual template uses at most four distinct letters.  The largest
difference between two block lengths is five, and each relevant legal prefix
also has at most four internally distinct symbols.  Hence all internal block
and prefix patterns are already realized over a four-symbol alphabet.  The
relative-label enumeration just described nevertheless permits the complete
configuration to use more than four symbols, and so it covers every joint
equality pattern occurring for any $q\geqslant4$.

The resulting finite certificate consists of $6\,445$ same-length
equality-pattern cases and $3\,709$ unequal-length cases.
Of the latter, $3\,613$ are accepted at the first stage and $96$ proceed to
the second stage, where $13\,469$ continuation equality-pattern cases are checked.
None of the tests fails.
The accompanying supplementary verifier reproduces this enumeration using
only the Python standard library.%
\footnote{A self-contained verifier, its documentation, and its expected output are available as ancillary material at the arXiv webpage of this article: 
https://doi.org/10.48550/arXiv.2509.06692.}
This proves the zero-error claim for every $q\geqslant4$.
For $q=2,3$, the family $\mathcal P_q$ is the restriction of
$\mathcal P_4$ to a smaller subalphabet, so the claim follows by heredity.

It remains to compute the rate.
The numbers $p_\ell$ of blocks of each length are
\begin{equation}
\label{eq:uniform-profile}
\begin{aligned}
&p_3=q, \qquad
p_4=(q)_2, \qquad
p_6=(q)_2+2(q)_3+(q)_4,  \\
&p_7=(q)_3, \qquad
p_8=3(q)_3+6(q)_4 .
\end{aligned}
\end{equation}
Since the criterion implies prefix-freeness, distinct formal concatenations
give distinct words, and
\begin{align}
  |\D_q(n)|=\sum_{\ell\in\{3,4,6,7,8\}}p_\ell|\D_q(n-\ell)|,
\end{align}
with $|\D_q(0)|=1$ and $|\D_q(n)|=0$ for $n<0$.
Because $p_3,p_4>0$, this recurrence is aperiodic, and standard generating
function arguments give $|\D_q(n)|=\Theta(\lambda_q^n)$.
The characteristic equation is precisely \eqref{eq:lambdaq}, which proves
\eqref{eq:Rinftynew}.
\qed
\end{proof}

For reference, the rates $\log_2\lambda_q$ for the smallest alphabets are
\begin{equation}
\label{eq:uniform-rates}
\begin{array}{c|ccccccc}
q&2&3&4&5&6&7&8\\ \hline
\log_2\lambda_q
&0.642805&1.044425&1.346292&1.571783&1.751473&1.901153&2.029649
\end{array}.
\end{equation}
Among the small alphabets compared explicitly here, the constructions from
\cite{benshimon+zabokritskiy,benshimon+zabokritskiy2} remain stronger for
$q=2,3$, whereas the present family improves the corresponding explicit rates
for $q=4,5$.
For $q=4$, their eleven-template baseline gives $1.324774\squeezeddots$, while
\eqref{eq:uniform-templates} gives $1.346292\squeezeddots$.
For $q=5$, their refined construction gives $1.554513\squeezeddots$, while ours
gives $1.571783\squeezeddots$.
For growing $q$, however, the run-wise lifting construction of
\cite{benshimon+zabokritskiy2}, which yields
$R_q^\star(1/2)\geqslant\log_2q-\log_2(1.82560995)+o(1)$, is asymptotically
stronger than the present bounded-template family.
The uniform family also improves the balanced two-class bound
\eqref{eq:Rinfty} for $3\leqslant q\leqslant8$; from $q=9$ onward,
\eqref{eq:Rinfty} is larger.
This crossover is consistent with the large-$q$ scalings of the two
particular constructions: \eqref{eq:lambdaq} gives
$\log_2\lambda_q=\frac{2}{3}\log_2q+o(\log_2 q)$ for the present template
family, whereas \eqref{eq:Rinfty} is $\log_2q-1+o(1)$.
The leading expressions balance at $q=8$; exactly, our bound remains larger
at $q=8$, while at $q=9$ it is $2.142361\squeezeddots$, below
$\log_2\sqrt{20}=2.160964\squeezeddots$.

The construction above was designed to be uniform in the alphabet size and was
not optimized separately for any particular value of $q$.
For any fixed alphabet, one may instead enumerate templates of bounded lengths,
test them using the finite local criterion, and search for large mutually
compatible subfamilies.
Such $q$-specific searches may yield further improvements, but we do not pursue
this computational optimization here.

Combining Theorems \ref{thm:GV} and \ref{thm:Rinfty} with
\eqref{eq:Rinfty} and \eqref{eq:joint}, we obtain the following bound, valid for every $q\geqslant2$ and
$\tau\in[0,\frac12]$,
\begin{align}
\label{eq:joint2}
  R_q^\star(\tau)
  \geqslant
  \max\left\{
  R_q^{\textsc{gv}}(\tau), \log_2\lambda_q, 
  \log_2\!\sqrt{\lfloor q/2\rfloor\!\cdot\!\lceil q/2\rceil}
  \right\}.
\end{align}

\subsection{A run-based upper bound}
\label{sec:run-upper}

We next derive an upper bound that depends nontrivially on the allowed
fraction of transpositions.  For $0\leqslant\mu\leqslant1-q^{-1}$, put
\begin{align}
\label{eq:Aq}
 A_q(\mu)
 &\defeq \log_2(q)H_q(\mu)
  =H(\mu)+\mu\log_2(q-1).
\end{align}
For $0<\mu\leqslant1-q^{-1}$ and
$0\leqslant\tau\leqslant\frac12$, define
\begin{align}
\label{eq:Gamma}
 \Gamma(\mu,\tau)
 &\defeq
 \max_{0\leqslant\theta\leqslant\min\{\tau, \mu/2\}}
 (\mu-\theta)
 H\!\left(\frac{\theta}{\mu-\theta}\right).
\end{align}
The expression in \eqref{eq:Gamma} is interpreted by continuity at the
endpoints.

\begin{theorem}
\label{thm:run-upper}
For every $q\geqslant2$ and $\tau\in[0,\frac12]$,
\begin{align}
\label{eq:run-upper}
 R_q^\star(\tau)
 &\leqslant U_q(\tau)
 \defeq
 \inf_{0<\mu\leqslant1-q^{-1}}
 \max\big\{A_q(\mu),\,\log_2(q)-\Gamma(\mu,\tau)\big\}.
\end{align}
\end{theorem}
\begin{proof}
Fix $0<\mu\leqslant1-q^{-1}$, let
$t_n=\lfloor\tau n\rfloor$, and split an arbitrary code
$\C_n\subseteq\Z_q^n$ correcting $t_n$ transpositions according to its
number of runs:
\begin{align}
\label{eq:run-split}
 \C_n^{\leqslant\mu}
 \defeq\{\myx\in\C_n \colon \run(\myx)\leqslant\mu n\},
 \qquad
 \C_n^{>\mu}
 \defeq\C_n\setminus\C_n^{\leqslant\mu}.
\end{align}
Lemma~\ref{thm:rho} gives
\begin{align}
\label{eq:low-run-code}
 |\C_n^{\leqslant\mu}|
 &\leqslant 2^{n A_q(\mu)+o(n)}.
\end{align}

Fix $0\leqslant\theta<\min\{\tau,\mu/2\}$ and choose integers $r_n$
such that $r_n/n\to\theta$ and $0\leqslant r_n\leqslant t_n$.
%The endpoints follow by continuity. 
If $\myx\in\C_n^{>\mu}$,
Proposition~\ref{thm:cardB}, the monotonicity of
$\binom{m-r_n}{r_n}$ in $m$, and
$\Beq(\myx;r_n)\subseteq B(\myx;t_n)$ imply
\begin{align}
 |B(\myx;t_n)|
 &\geqslant \binom{\run(\myx)-r_n}{r_n}
 \geqslant \binom{\lfloor\mu n\rfloor-r_n}{r_n} \notag\\
 &\geqslant
 2^{n (\mu-\theta)
 H\left(\frac{\theta}{\mu-\theta}\right)-o(n)}.
 \label{eq:high-run-ball}
\end{align}
The last inequality follows from Stirling's approximation and is uniform in
$\myx\in\C_n^{>\mu}$.  Maximizing over $\theta$ and using continuity at
the boundary therefore gives
\begin{align}
\label{eq:high-run-ball-optimized}
 |B(\myx;t_n)|
 &\geqslant 2^{n \Gamma(\mu,\tau)-o(n)}.
\end{align}
For $\tau=0$, the same statement holds with $r_n=0$ and
$\Gamma(\mu,0)=0$.
The balls centered at distinct codewords are disjoint.  Hence
\begin{align}
\label{eq:high-run-packing}
 |\C_n^{>\mu}|
 \leqslant
 2^{n\left(\log_2(q)-\Gamma(\mu,\tau)\right)+o(n)}.
\end{align}
Combining \eqref{eq:low-run-code} and \eqref{eq:high-run-packing}, applying
the result to an optimal code, taking the asymptotic rate, and finally taking
the infimum over $\mu$ proves \eqref{eq:run-upper}.
\qed
\end{proof}

At $\tau=0$, one has $\Gamma(\mu,0)=0$, and hence
$U_q(0)=\log_2(q)$, as expected.

The optimization in Theorem~\ref{thm:run-upper} has a simple explicit
form.  Put
\begin{align}
\label{eq:golden-parameters}
 \varphi
 &\defeq\frac{1+\sqrt5}{2},
 \\
 \rho^\dagger
 &\defeq\frac{5-\sqrt5}{10},
\end{align}
and, for $0\leqslant\rho\leqslant\frac12$, define
\begin{align}
\label{eq:h-rho}
 h(\rho)
 &\defeq(1-\rho)
 H\!\left(\frac{\rho}{1-\rho}\right).
\end{align}
For $0<\rho<\frac12$, direct differentiation gives
\begin{align}
\label{eq:h-derivative}
 h'(\rho)
 &=\log_2\!\left(\frac{(1-2\rho)^2}{\rho(1-\rho)}\right).
\end{align}
It follows that $h$ has its unique maximum at $\rho^\dagger$ and that
\begin{align}
\label{eq:h-maximum}
 h(\rho^\dagger)
 &=\log_2\varphi.
\end{align}
The change of variables $\rho=\theta/\mu$ in \eqref{eq:Gamma} therefore
gives
\begin{align}
\label{eq:Gamma-reduced}
 \Gamma(\mu,\tau)
 &=\mu\,
 h\!\left(\min\left\{\frac{\tau}{\mu},\rho^\dagger\right\}\right).
\end{align}

This also gives the saturated value of the bound for every alphabet size.
Let $\mu_q\in(0,1-q^{-1})$ be the unique solution of
\begin{align}
\label{eq:muq}
 H(\mu_q)+\mu_q\log_2(q-1)
 &=\log_2(q)-\mu_q\log_2\varphi.
\end{align}
Then
\begin{align}
\label{eq:general-saturation}
 U_q(\tau)
 &=U_q^{\mathrm{sat}}
 \defeq H(\mu_q)+\mu_q\log_2(q-1)
  =\log_2(q)-\mu_q\log_2\varphi,
 &
 \tau&\geqslant\rho^\dagger\mu_q.
\end{align}
Indeed, \eqref{eq:h-maximum} implies
$\Gamma(\mu,\tau)\leqslant\mu\log_2\varphi$ for every $\tau$.
The first side of \eqref{eq:muq} is strictly increasing on
$[0,1-q^{-1}]$, while the second is strictly decreasing, so their
intersection gives a lower bound on the maximum in \eqref{eq:run-upper} for
every $\mu$.  If $\tau\geqslant\rho^\dagger\mu_q$, equality in
\eqref{eq:h-maximum} is attained at $\mu=\mu_q$, proving
\eqref{eq:general-saturation}.  In particular, $\mu_q\to1$ as
$q\to\infty$, and hence
\begin{align}
\label{eq:general-saturation-asymptotic}
 U_q^{\mathrm{sat}}
 &=\log_2q-\log_2\varphi+o(1)
 \qquad (q\to\infty).
\end{align}
At $\tau=1/2$, \eqref{eq:general-saturation-asymptotic} recovers the
large-alphabet zero-error upper bound of
\cite[Theorem~29]{benshimon+zabokritskiy2}.  The contribution of
Theorem~\ref{thm:run-upper} is the finite-$q$ converse for every
$\tau\in[0,\frac12]$, including its nonconstant, $\tau$-dependent part before
saturation.

\begin{figure}[h]
 \centering
  \includegraphics[width=0.95\columnwidth]{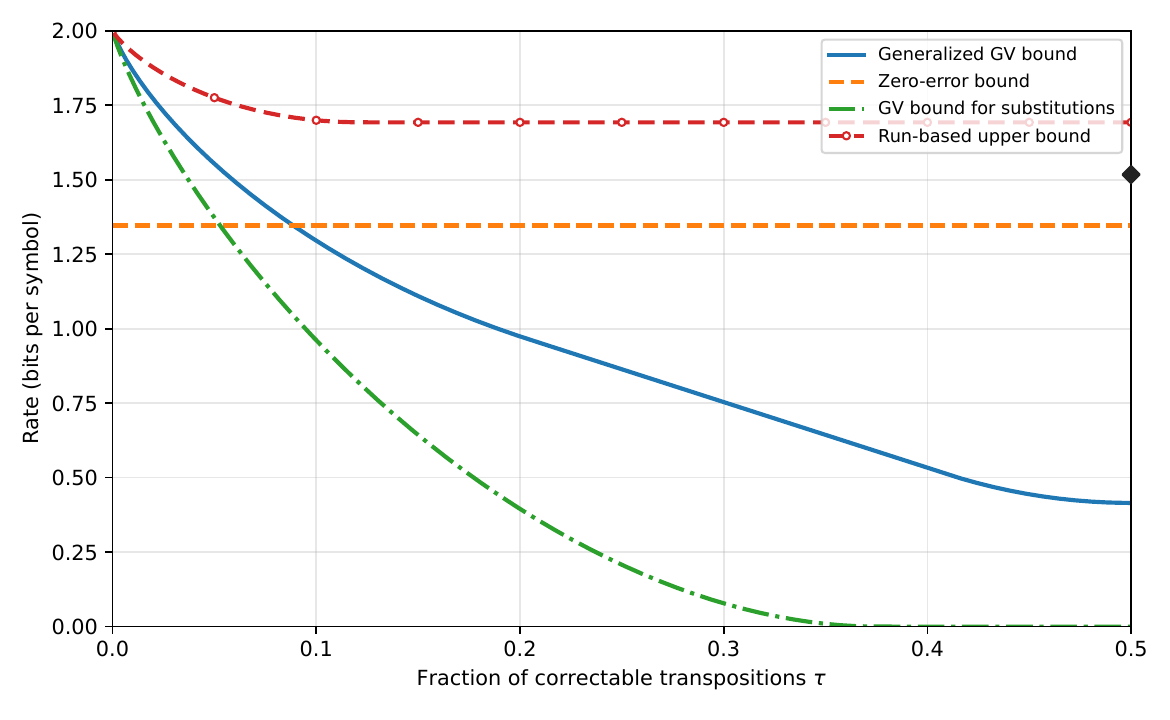}
\caption{Lower and upper bounds, in bits per symbol, on the asymptotic rate
of codes correcting a fraction $\tau$ of pairwise disjoint adjacent
transpositions over an alphabet of size $q=4$.
The lower bounds are the generalized GV bound from Theorem~\ref{thm:GV}, the
substitution-derived GV bound \eqref{eq:GVsub}, and the zero-error bound
$\log_2\lambda_4=1.346292\squeezeddots$ from
Theorem~\ref{thm:Rinfty}.  The zero-error lower bound exceeds the generalized
GV bound for $\tau>0.088654\squeezeddots$. The upper bound is the one from Theorem~\ref{thm:run-upper}.
The isolated point at $\tau=1/2$ is the zero-error upper bound
$R_4^\star(1/2)\leqslant\frac{1}{18}\log_2(550^3-24)
=1.517214\squeezeddots$ from
\cite[Theorem~28]{benshimon+zabokritskiy2}.}
\label{fig:GV}
\end{figure}

For $q=4$, one has $\mu_4=0.442887\squeezeddots$, and
\begin{align}
\label{eq:q4-saturation}
 U_4(\tau)
 =1.692528\squeezeddots,
 \qquad
 \tau\geqslant\rho^\dagger\mu_4
 =0.122411\squeezeddots.
\end{align}
For example, $U_4(0.05)=1.775357\squeezeddots$.
The flat portion of this curve reflects saturation of the exponential
ball-size estimate supplied by Proposition~\ref{thm:cardB}, rather than a
claim that the true rate is constant.  At the zero-error endpoint, the
covering construction in
\cite[Theorem~28]{benshimon+zabokritskiy2} gives the stronger bound
\begin{align}
\label{eq:q4-zero-error-upper}
 R_4^\star(1/2)
 &\leqslant\frac{1}{18}\log_2(550^3-24)
 =1.517214\squeezeddots.
\end{align}
We indicate this endpoint bound separately in Figure~\ref{fig:GV}; it is not
part of the run-based curve \eqref{eq:run-upper}.

Figure~\ref{fig:GV} compares the run-based upper bound with the generalized
GV bound, the substitution-derived GV bound \eqref{eq:GVsub}, and the
zero-error lower bound for $q=4$.

\section{Bounds on codes correcting $ t = \textnormal{const} $ transpositions}
\label{sec:codet}

In this section we present bounds on the cardinality of optimal codes in $ \Z_q^n $
correcting $ t = \textnormal{const} $ transpositions.
In contrast with the linear $ t = \tau n $ regime, the lower bound obtained via the
Gilbert--Varshamov argument is not better than the bounds that can be obtained by
different methods, see Remarks \ref{rem:binary} and \ref{rem:subcodes} below.
Nonetheless, we derive the expression for the average ball size and state the
corresponding bound in the $ t = \textnormal{const} $ regime as they may be of
separate interest.

Lemma \ref{thm:rho} implies that, for any fixed $ 0 < \epsilon < 1 - q^{-1} $, the
number of strings having at most $ (1 - q^{-1} - \epsilon) n $ runs is \emph{exponentially}
smaller than the number of remaining strings, namely
\begin{align}
\label{eq:runexp}
  \left| \{\myx \in \Z_q^n \colon \run(\myx) \leqslant (1 - q^{-1} - \epsilon) n \} \right|
	=  q^{n H_q(1 - q^{-1} - \epsilon) + o(n)} ,
\end{align}
where $ H_q(1 - q^{-1} - \epsilon) < 1 $.
A very useful consequence of this fact is that one can disregard such strings
in the asymptotic analysis of optimal codes in the $ n \to \infty $,
$ t = \textnormal{const} $ regime.
Namely, since $ M^\star_q(n; t) $ is lower-bounded by the cardinality of optimal
codes correcting $ t $ substitutions (see Section \ref{sec:subvstransp}), and
the latter for constant $ t $ scales as $ \Omega(\frac{q^{n}}{n^{2t}}) $ (by
the standard Gilbert--Varshamov bound in the Hamming metric \cite[Section 2.8]{huffman+pless}),
we conclude that
\begin{align}
\label{eq:Ttheta}
  M^\star_q(n; t) = q^{n-o(n)} .
\end{align}
Relations \eqref{eq:runexp} and \eqref{eq:Ttheta} imply that, for an optimal
code correcting $ t $ transpositions, throwing out its codewords that have at
most $ (1 - q^{-1} - \epsilon) n $ runs affects its cardinality by an
asymptotically negligible amount.

\begin{theorem}
\label{thm:bounds}
For any fixed $ q \geqslant 2 $ and $ t \geqslant 1 $, as $ n \to \infty $,
\begin{align}
\label{eq:upboundadj}
  \frac{(2t)! \, q^{2t}}{(q-1)^{2t}} \frac{q^{n}}{n^{2t}}
  \;\lesssim\;
	M^\star_q(n; t)
	\;\lesssim\;
	\frac{t! \, q^{t}}{(q-1)^{t}} \frac{q^{n}}{n^t}  .
\end{align}
%In the binary case ($ q = 2 $), the lower bound can be improved to $ c_t \frac{2^{n+t}}{n^t} $,
%where $ c_1 = \frac{1}{2} $, $ c_2 = \frac{1}{3} $, and $ c_t = \frac{1}{2t+1} $
%for $ t \geqslant 3 $.
\end{theorem}
\begin{proof}
From Proposition \ref{thm:Teq} we can find the asymptotic behavior of $ \Teq_q(n;r) $
for fixed $ r $ and $ n \to \infty $,
\begin{align}
\label{eq:Teqr}
  \Teq_q(n;r)  \sim  n^r \frac{1}{r!} q^{n-r} (q-1)^r ,
\end{align}
where we used the fact that $ \binom{n}{s} \sim \frac{n^s}{s!} $ when $ s $ is
fixed and $ n \to \infty $.
From \eqref{eq:Teqr}, \eqref{eq:PT} and \eqref{eq:TTbar} it follows that
%$ T_q(n;r) $ and $ \bar{T}_q(n;r) $ have the same asymptotic scaling as $ \Teq_q(n;r) $
\begin{align}
  \bar{T}_q(n;r)  \sim  T_q(n;r)  \sim  \Teq_q(n;r)
\end{align}
in the regime currently being analyzed.
This fact, together with \eqref{eq:GV1}, implies the left-hand side of
\eqref{eq:upboundadj}.

In order to show the right-hand inequality, an asymptotic lower bound on the
cardinality of $ B(\myx; t) $ (the set of all strings that can be obtained by
applying at most $ t $ transpositions to the string $ \myx $) will be needed.
By using Proposition \ref{thm:cardB}, for strings satisfying $ \run(\myx) > (1 - q^{-1} - \epsilon)n $,
we get
\begin{align}
|B(\myx; t)| \geqslant |\Beq(\myx; t)|
\geqslant \binom{\lfloor(1 - q^{-1} - \epsilon)n\rfloor - t}{t}
\label{eq:tmp}
\sim  \frac{(1 - q^{-1} - \epsilon)^t \, n^t}{t!} .
\end{align}
We can now derive the upper bound in \eqref{eq:upboundadj} by a packing argument.
Let $ \C^\star_q(n; t) $ be an optimal code correcting $ t $ transpositions,
meaning that $ |\C^\star_q(n; t)| = M^\star_q(n; t) $.
By \eqref{eq:disjoint}, the sets $ B(\myx; t) $ centered at codewords of
$ \C^\star_q(n; t) $ do not overlap, so we have
\begin{align}
\label{eq:upper1}
  \sum_{\myx \in \C^\star_q(n;t) \colon \run(\myx) > (1-q^{-1}-\epsilon)n} |B(\myx; t)|
	\leqslant
	\sum_{\myx \in \C^\star_q(n;t)} |B(\myx; t)|
	\leqslant
	|\Z_q^n|  =  q^n .
\end{align}
Together with \eqref{eq:tmp} this implies that
\begin{align}
\label{eq:upper2}
  \frac{1}{t!} \left(1 - q^{-1} - \epsilon\right)^{\! t} n^t \cdot \left| \left\{\myx \in \C^\star_q(n;t) \colon \run(\myx) > (1-q^{-1}-\epsilon)n \right\} \right|
	\lesssim	q^n ,
\end{align}
From Lemma~\ref{thm:rho} and the discussion preceding this theorem we know that we can safely
ignore in the asymptotic analysis the codewords with at most $ (1 - q^{-1} - \epsilon) n $
runs, meaning that
\begin{align}
\label{eq:upper2p}
  \left| \left\{\myx \in \C^\star_q(n;t) \colon \run(\myx) > (1-q^{-1}-\epsilon)n \right\} \right|
  \sim  \left| \C^\star_q(n;t) \right|  =  M^\star_q(n; t).
\end{align}
Now \eqref{eq:upper2} and \eqref{eq:upper2p} imply
\begin{align}
\label{eq:upper3}
  \frac{1}{t!} \left(1 - q^{-1} - \epsilon\right)^{\! t} n^t  \cdot  M^\star_q(n; t)  \lesssim  q^n .
\end{align}
Since \eqref{eq:upper3} holds for any $ \epsilon > 0 $, the upper bound in
\eqref{eq:upboundadj} is established.
\qed
\end{proof}

\begin{remark}[Lower bound for the binary case]
\label{rem:binary}
\textnormal{
For $ q = 2 $, a lower bound better than the one in Theorem \ref{thm:bounds}
was derived in \cite[Theorem 11]{kovacevic} by using an interpretation of binary
codes correcting \emph{successive} transposition errors as packings
in the Manhattan metric, and using the densest known packings, namely
\begin{align}
  M^\star_2(n; t)  \gtrsim  c_t \frac{2^{n+t}}{n^t} ,
\end{align}
where $ c_1 = \frac{1}{2} $, $ c_2 = \frac{1}{3} $,%
\footnote{\cite[Theorem 11]{kovacevic} states a slightly worse constant
$ c_2 = \frac{1}{4} $.
It can be shown based on \cite[Theorem 3]{xiao+zhou} that $ c_2 $ can be improved
to $ \frac{1}{3} $.}
and $ c_t = \frac{1}{2t+1} $ for $ t \geqslant 3 $.
In particular, in the binary case we have
\begin{align}
\label{eq:M2}
  M^\star_2(n; t) = \Theta\Big(\frac{2^n}{n^t}\Big) .
\end{align}
Unfortunately, as already mentioned in Remark \ref{rem:binarylin}, this geometric
interpretation fails when $ q > 2 $ and thus the construction that implies this
lower bound does not generalize to larger alphabets.
\myqed}
\end{remark}

\begin{remark}[Lower bound via substitution-correcting codes]
\label{rem:subcodes}
\textnormal{
For general $ q $, a lower bound better than the one in \eqref{eq:upboundadj}
follows from the known bounds for substitution-correcting codes (see Section
\ref{sec:subvstransp}).
For example, it is known that $ q $-ary codes correcting a single substitution
and having cardinality $ \Theta(\frac{q^n}{n}) $ exist \cite{huffman+pless}
(e.g., Hamming codes), and hence
\begin{align}
\label{eq:T1}
  M^\star_q(n; 1)  =  \Theta\Big(\frac{q^n}{n}\Big) .
\end{align}
For $ t > 1 $, Varshamov's improvement of Gilbert's bound \cite[Section 2.9]{huffman+pless}
implies that
\begin{align}
  M^\star_q(n; t)  =  \Omega\Big(\frac{q^n}{n^{2t-1}}\Big) .
\end{align}
Further improvements of the exponent in the denominator are possible in some cases;
see, e.g., \cite{dumer,yekhanin+dumer}.
}

\textnormal{
Based on what is known in the special case $ t = 1 $ (see \eqref{eq:T1}),
as well as in the case of general $ t $ and $ q = 2 $ (see \eqref{eq:M2}),
it is reasonable to conjecture that
\begin{align}
\label{eq:conj}
  M^\star_q(n; t)  \stackrel{?}{=}  \Theta\Big(\frac{q^n}{n^t}\Big) ,
\end{align}
i.e., that the redundancy of $ t \log_q n + {\mathcal O}(1) $ is achievable for
any fixed $ t $ and alphabet size $ q $.
At present, however, we do not have a proof of this fact.
\myqed}
\end{remark}

\section{Conclusion and further work}
\label{sec:conclusion}

We have analyzed the problem of correcting transpositions of consecutive symbols
in $ q $-ary strings---types of errors that may occur in various contexts, e.g.,
as synchronization and timing errors, spelling errors, inter-symbol interference
errors, genome rearrangement errors, etc.

In the linear regime $ t = \tau n $, we have presented a generalized
Gilbert--Varshamov lower bound on the achievable code rates.
For the zero-error case, we constructed an alphabet-uniform family of
symbolic templates certified by a finite local criterion, improving the
available $q=4,5$ bounds.
Since every zero-error code also corrects any prescribed fraction of
transpositions, this construction strengthens the achievable-rate envelope
beyond the GV bound for sufficiently large $\tau$.
We also derived a run-based upper bound for the full linear regime.  The bound
depends nontrivially on $\tau$, although it saturates before the zero-error
endpoint because the exponential estimate obtained from the conflict graph of
run boundaries saturates.  Sharpening this converse and finding efficiently encodable and
decodable code families approaching the generalized GV bound remain open problems.
For codes of infinite minimum distance, the problem of deriving the exact value
of the maximum achievable rate, i.e., the zero-error capacity, is open for all
alphabet sizes.
In the $ t = \textnormal{const} $ regime, as one of the main open questions in
this line of work we mention a construction of nonbinary codes correcting $ t > 1 $
transpositions whose cardinality scales as $ \Theta(\frac{q^n}{n^t}) $ when
$ n \to \infty $ (see \eqref{eq:conj}).

Finally, there are several related models describing communication in the presence
of synchronization and reordering errors that are potentially relevant for modern
information transmission and storage systems and that are not yet fully understood,
e.g., timing channels, permutation channels with bounded displacements, channels
with a combination of transpositions and other types of errors, etc.

\vspace{5mm}
{\small
\subsubsection{\small Acknowledgments.}
%M. Kova\v{c}evi\'{c} would like to dedicate this article to the students and
%teachers of Serbia who stood bravely in the face of corruption and injustice
%during the academic year 2024/2025. 
M. Kova\v{c}evi\'{c} was supported by the Ministry of Science, Technological Development
and Innovation of the Republic of Serbia (contract no. 451-03-34/2026-03/200156)
and by the Faculty of Technical Sciences, University of Novi Sad, Serbia (project
no. 01-3609/1).

%H. M. Kiah was supported by...

K. Goyal was supported by the French government through the France 2030 investment plan managed by the National Research Agency (ANR), as part of the Initiative of
Excellence Université Côte d’Azur under reference number ANR-15-IDEX-01, and by the ANR grant ANR-22-CPJ2-0054-01.

%\subsubsection{\small Disclosure of Interests.}
%The authors have no competing interests to declare that are relevant to the content
%of this article.
%The authors state that they do not have any conflicts of interest concerning the
%publication of this article.
}

\end{document}